\documentclass[prl,twocolumn,letterpaper,superscriptaddress]{revtex4-1}
\usepackage[space]{grffile}
\usepackage{bm,graphicx,graphics,amsmath,amssymb,bm,epsfig,color}
\usepackage{euscript,tabularx}
\usepackage{longtable}
\usepackage{float}
\usepackage[draft]{todonotes}

\graphicspath{{./figs/final/}}

\begin{document}

\title{Spin conductance in extended thin films of YIG driven from
  thermal to subthermal magnons regime by large spin-orbit torque}

\author{N. Thiery}
\affiliation{SPINTEC, CEA-Grenoble, CNRS and Universit\'e Grenoble Alpes, 38054
  Grenoble, France}

\author{A. Draveny}
\affiliation{SPINTEC, CEA-Grenoble, CNRS and Universit\'e Grenoble Alpes, 38054
  Grenoble, France}

\author{V. V. Naletov} 
\affiliation{SPINTEC, CEA-Grenoble, CNRS and Universit\'e Grenoble Alpes, 38054
  Grenoble, France}
\affiliation{Institute of Physics, Kazan Federal University, Kazan
    420008, Russian Federation}

\author{L. Vila}
\affiliation{SPINTEC, CEA-Grenoble, CNRS and Universit\'e Grenoble Alpes, 38054
  Grenoble, France}

\author{J.P. Attan\'e}
\affiliation{SPINTEC, CEA-Grenoble, CNRS and Universit\'e Grenoble Alpes, 38054
  Grenoble, France}

\author{C. Beign\'e}
\affiliation{SPINTEC, CEA-Grenoble, CNRS and Universit\'e Grenoble Alpes, 38054
  Grenoble, France}

\author{G. de Loubens} 
\affiliation{SPEC, CEA-Saclay, CNRS, Universit\'e Paris-Saclay,
  91191 Gif-sur-Yvette, France}

\author{M. Viret}
\affiliation{SPEC, CEA-Saclay, CNRS, Universit\'e Paris-Saclay,
  91191 Gif-sur-Yvette, France}

\author{N. Beaulieu} 
\affiliation{SPEC, CEA-Saclay, CNRS, Universit\'e Paris-Saclay,
  91191 Gif-sur-Yvette, France}
\affiliation{LabSTICC, CNRS, Universit\'e de Bretagne Occidentale,
  29238 Brest, France}

\author{J. Ben Youssef} 
\affiliation{LabSTICC, CNRS, Universit\'e de Bretagne Occidentale,
  29238 Brest, France}

\author{V. E. Demidov}
\affiliation{Department of Physics, University of Muenster, 48149 Muenster, Germany}

\author{S. O. Demokritov} 
\affiliation{Department of Physics, University of Muenster, 48149 Muenster, Germany}
\affiliation{Institute of Metal Physics, Ural Division of RAS,
  Yekaterinburg 620041, Russian Federation}

\author{A.~N. Slavin} \affiliation{Department of Physics, Oakland
  University, Michigan 48309, USA}

\author{V.~S. Tiberkevich} \affiliation{Department of Physics, Oakland
  University, Michigan 48309, USA}

\author{A. Anane} 
\affiliation{Unit\'e Mixte de Physique CNRS, Thales, Universit\'e
  Paris-Saclay, 91767 Palaiseau, France}

\author{P. Bortolotti} 
\affiliation{Unit\'e Mixte de Physique CNRS, Thales, Universit\'e
  Paris-Saclay, 91767 Palaiseau, France}

\author{V. Cros}
\affiliation{Unit\'e Mixte de Physique CNRS, Thales, Universit\'e
  Paris-Saclay, 91767 Palaiseau, France}

\author{O. Klein}
\email[Corresponding author:]{ oklein@cea.fr}
\affiliation{SPINTEC, CEA-Grenoble, CNRS and Universit\'e Grenoble Alpes, 38054
  Grenoble, France}

\date{\today}

\begin{abstract}
  We report a study on spin conductance through ultra-thin extended
  films of epitaxial Yttrium Iron Garnet (YIG), where spin transport
  is provided by propagating spin waves, that are generated and
  detected by direct and inverse spin Hall effects in two Pt wires
  deposited on top. While at low current the spin conductance is
  dominated by transport of exchange magnons, at high current, the
  spin conductance is dominated by magnetostatic magnons, which are
  low-damping non-equilibrium magnons thermalized near the spectral
  bottom by magnon-magnon interaction, with consequent a sensitivity
  to the applied magnetic field. This picture is supported by
  microfocus Brillouin Light Scattering spectroscopy.
\end{abstract}

\maketitle

\begin{figure}
  \includegraphics[width=8.5cm]{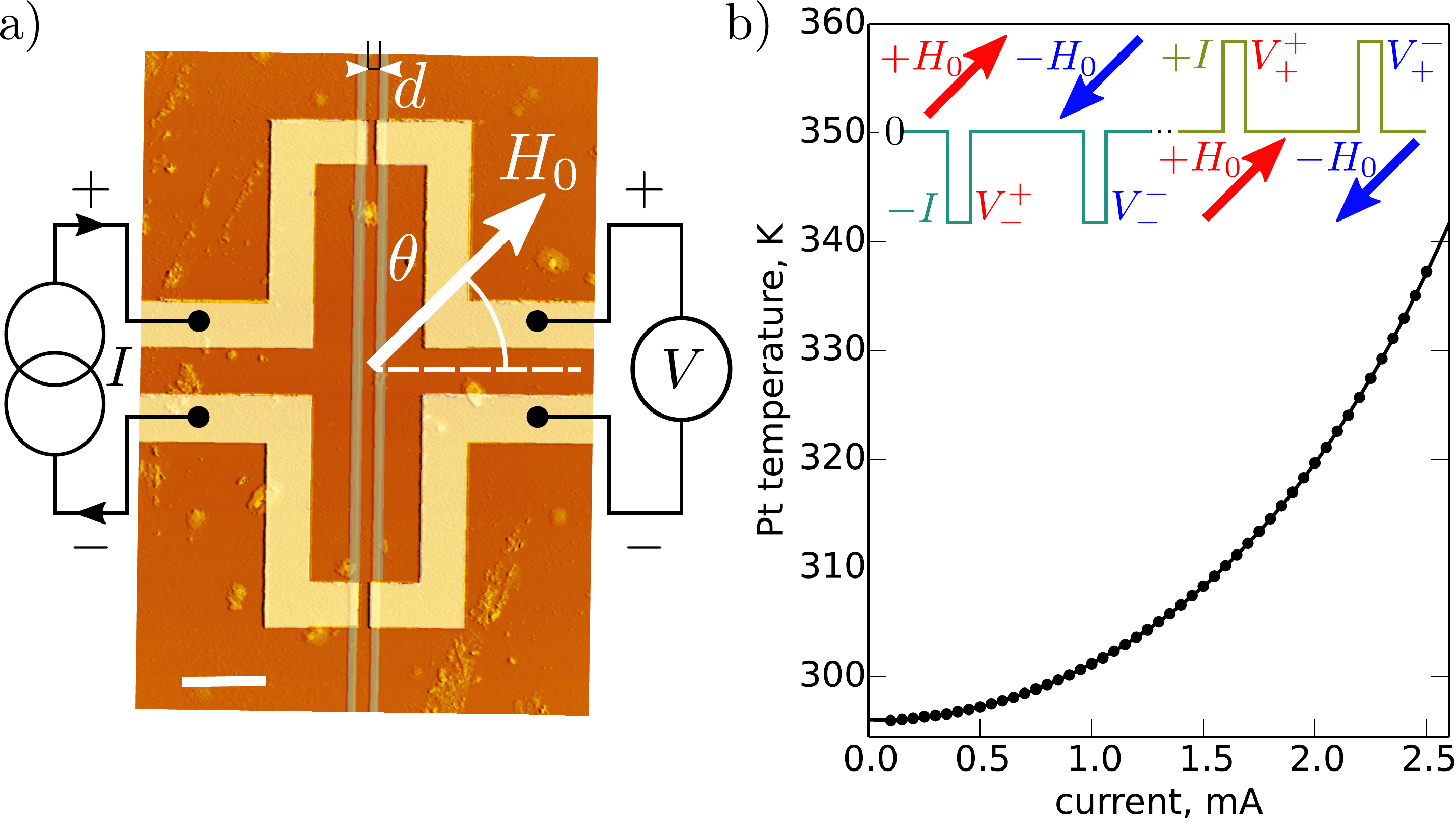}
  \caption{(Color online) a) Top view of the lateral device. Two
    vertical Pt wires (grey) are placed at a distance $d= 1.2$~$\mu$m
    apart on top of a 18~nm thick YIG film (scale bar is
    5~$\mu$m). The non-local conductance $I$-$V$ (injector-detector)
    is measured using negative and positive current pulses while
    rotating the magnetic field $H_0$ in-plane by an angle
    $\theta$. Panel b) shows the temperature elevation produced in the
    Pt injector by Joule heating while increasing the pulse amplitude
    $I$.}
  \label{FIG1}
\end{figure}

The recent demonstrations that spin orbit torques (SOT) allow one to
generate and detect pure spin currents \cite{valenzuela06, kajiwara10,
  miron11, rojas13a, sangiao15, mellnik14, lesne16, chauleau16} has
triggered a renewed effort to study magnons transport in extended
magnetic films. This topic is currently recognized as one of the
important emerging research direction in modern magnetism
\cite{Sander2017}. This is because, in contrast to spin transfer
process in confined geometries (\textit{e.g.} nano-pillars or
nano-contacts) where usually the uniform magnon mode dominate the
dynamics, very little is known about spin transfer in extended
geometries, which have continuous spin-wave spectra containing many
modes which can take part in the magnon-magnon interactions. A large
effort has concentrated so far on yttrium iron garnet (YIG), a
magnetic insulator, which is famous for having the lowest known
magnetic damping parameter \cite{spencer59}. From a purely fundamental
point of view, the studies of magnon transport in YIG by means of the
direct and inverse spin Hall effects (ISHE) \cite{kajiwara10, wang11,
  padron-hernandez11, chumak12, hahn13, kelly13, hamadeh14b, collet16,
  lauer16, Wesenberg2017} are very interesting, as they provide new
means to alter efficiently the energy distribution of magnons and,
potentially, even to trigger condensation \cite{Bender2012}.

Contrary to the case of magnons excited coherently, \textit{e.g.} by
means of a ferromagnetic resonance or parametric pumping, when the
frequencies of the excited magnons are fully determined by the
frequencies of the external signals, the excitation of magnons by
means of spin transfer process lacks frequency selectivity
\cite{demidov11d}, and, therefore, can lead to their excitation in a
broad frequency range. This poses a challenge for the identification
of the nature of magnons modes excited by SOT. It has been already
shown in \cite{Tikhonov2013}, that it is convenient and useful to
introduce the concepts of subthermal (having energy close to the
bottom of the spin wave spectrum) and thermal (having energy close to
$k_B T$) magnons. On one hand, it has been well established
\cite{demokritov:06, Demidov2007} that subthermal magnons can be very
efficiently thermalized near the spectral bottom (region of so-called
magnetostatic waves) by the intensive magnon-magnon interaction, whose
decay rate between quasi-degenerate modes increases with power, to
reach a quasi-equlibrium state by a non-zero chemical potential
\cite{demokritov:06, Demidov2007, Du2017} and an effective temperature
\cite{Serga2014}. On the other hand, it has been shown, both
experimentally and theoretically \cite{Serga2014}, that the groups of
subthermal and thermal magnon are effectively decoupled from each
other, as under the intensive parametric pumping one can reach a state
where the effective temperature of subthermal magnons exceeds the real
remperature characterizing the thermal magnons by a factor of 100.

Under spin transfer process, whose efficiency is known to increase
with decreasing magnon frequency, in confined geometries with
localized spin-current injection (\textit{i.e.} when there are no
quasi-degenerate modes), it has been shown, that one can reach current
induced coherent GHz-frequency magnon dynamics in YIG
\cite{hamadeh14b, collet16, demidov16a}). In extended thin-films, the
recently discovered non-local magnon transport \cite{cornelissen15,
  goennenwein15, li16, wu16, cornelissen16} suggests that the magnon
transport properties of YIG films subjected to small SOT are dominated
by thermal magnons, whose number overwhelmingly exceeds the number of
other modes at any non-zero temperature. The interesting challenge is
to elucidate what will happen to this spectrum (in particular the
interplay between the thermal and subthermal part \cite{Lendinez2017})
when one applies large SOT to a magnon continuum.

We propose herein to measure the room temperature spin conductance of
YIG films when the driving current is varied in a wide range
magnitudes \cite{flebus16a,bender16} creating, first, a
quasi-equilibrium transport regime, and, then, driving the system to a
strongly out-of-equilibrium state. To reach this goal the spin current
injected in the YIG by SOT shall be increased by more than one order
of magnitude compared to previous works, while simultaneously reducing
the film thickness by also an order of magnitude, using ultra-thin
films of YIG grown by liquid phase epitaxy (LPE)
\cite{castel12b,hahn13}. A series of lateral devices have been
patterned on a 18 nm thick YIG films. Ferromagnetic resonance (FMR)
characterization of the bare film are summarized in Table 1. On these
films, we have deposited Pt wires, 10~nm thick, 300~nm wide, and
20~$\mu$m long. The measured resistance of the Pt wire at room
temperature is $R_0=1.3$~k$\Omega$. The lateral device geometry is
shown in FIG.\ref{FIG1}a. One monitors the voltage $V$ along one wire
as a current $I$ flows through a second wire separated by a gap of
1.2~$\mu$m. Here the Pt wires are connected by 50~nm thick Al
electrodes colored in yellow. Since large amount of electrical current
needs to flow in the Pt, a pulse method is used to reduce
significantly Joule heating. In the following the current is injected
during 10~ms pulses series enclosed in a 10\% duty cycle. Temperature
sensing is provided by the change of relative resistance of the Pt
wire during the pulse.  In FIG.\ref{FIG1}b, we have plotted
$\kappa_\text{Pt} (R_I-R_0)/R_0$ as a function of the current $I$,
where the coefficient $\kappa_\text{Pt} = 254$~K is specific to Pt. We
observe that the pulse method allows to keep the absolute temperature
of our YIG below 340~K \footnote{YIG spontaneous magnetization
  decreases by about 4G/$^\circ$C} at the maximum current amplitude of
2.5~mA. Avoiding excessive heating of the YIG is crucial because, in a
joint review paper \cite{Thiery2017a}, it is shown that epitaxial YIG
films grown by LPE behave as a large gap semiconductor, with an
electrical resistivity that decreases exponentially with increasing
temperature following an activated behavior. As shown in
Ref.\cite{Thiery2017a}, at 340~K, however, the electrical resistivity
of YIG remains larger than $10^6$~$ \Omega\cdot\text{cm}$ and thus the
YIG can still be considered a good insulator ($R>30$~G$\Omega$) over
the current range explored herein.

\begin{table}
  \caption{Summary of the physical properties of the materials used in this study.}
  \begin{ruledtabular}
    \begin{tabular}{c | c c c c }
      YIG & $t_\text{YIG}$ (nm) & $4 \pi M_s$ (G) & {$\alpha_\text{YIG}$} &
      {$\Delta H_0$} (Oe)\\ \hline
      & 18 & $1.6 \times 10^3$ & {$4.4 \times 10^{-4}$}  & {3.7} \\ \hline \hline 
      Pt & $t_\text{Pt}$ (nm) & $\rho$ ($\mu \Omega$.cm) &
      $\alpha_\text{YIG$\mid$Pt}$ & $g_{\uparrow \downarrow}$ (m$^{-2}$) \\ \hline 
      & 10 & 19.5 & $2.4  \times 10^{-3}$ & $3 \times 10^{18}$ 
 \end{tabular}
\end{ruledtabular}\label{tab:mat}
\end{table}

\begin{figure}
  \includegraphics[width=8.5cm]{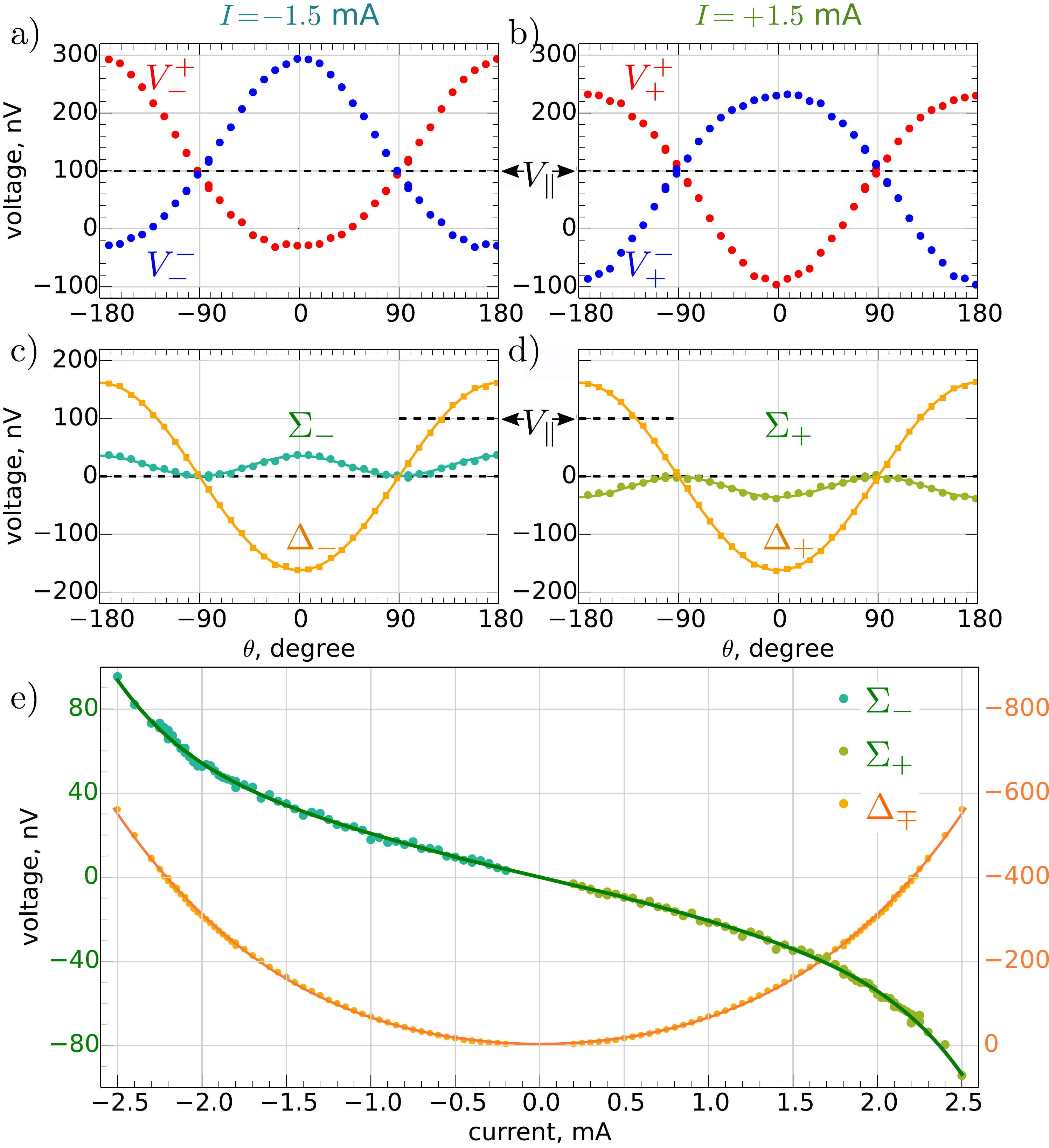}
  \caption{(Color online) Angular dependence of the non-local voltages
    $V_\mp^\pm$ measured while inverting the polarity of the applied
    field $H_0=\pm 2\text{kOe}$ (red/blue) respectively for a)
    negative and b) positive current pulses $I=\mp 1.5 \text{mA}$. The
    measured signal can be decomposed c) and d) in three components:
    $\Sigma$ (green): the signal sum, $\Delta$ (orange): the signal
    difference and $V_{\parallel}$: the offset; respectively even/odd,
    odd/even in field/current, and an independent contribution
    (dashed). Panel e) shows the current dependence of the amplitude
    $\Sigma$ and $\Delta$.}
  \label{FIG2}
\end{figure}

The lateral device is biased by an in-plane magnetic field, $H_0$ set
at a variable polar angle $\theta$, with respect to the perpendicular
of the Pt wires. FIG.\ref{FIG2}a-d displays the results when
$I=1.5$~mA and $H_0=2$~kOe \footnote{$H_0$ is much higher than the
  Oersted field of the current ($< 50$~Oe)}. For each value of
$\theta$, 4 measurements $V^\pm_\mp$ are performed corresponding to
the 4 combinations of the polarities of $\pm H_0$ and $\mp I$ (the
polarity convention is defined in FIG.\ref{FIG1}a). FIG.\ref{FIG2}a
and \ref{FIG2}b show the raw data obtained respectively for negative
and positive current pulses. Clearly the non-local voltage oscillates
around an offset, $V_\parallel$, defined as the voltage measured at
$\theta = \pm 90^\circ$. This offset is independent of the current
polarity and its amplitude scales with the temperature elevation of Pt
produced by Joule heating (see FIG.2c in \cite{Thiery2017a}). We
ascribe it to thermoelectric effects produced by a small temperature
difference at the two Pt$|$Al contacts of the detector circuit
\footnote{Considering the Seebeck coefficient of Pt$|$Al, the
  100~$\mu$Voffset measured in FIG.\ref{FIG1}b, corresponds to a
  temperature difference of less than 0.03~$^\circ$C between the top
  and bottom electrodes.}. By contrast, the anisotropic part of the
voltage is ascribed to magnons transport.

To gain more insight, the data are sorted according to their symmetry
with respect to the $(H_0,I)$-polarity. This is done in
FIG.\ref{FIG2}c and \ref{FIG2}d by constructing the signal sum $\Sigma_\mp =
(V^+_\mp+V^-_\mp)/2-V_\parallel$ (green tone) even in field and the
signal difference $\Delta_\mp = (V^+_\mp-V^-_\mp)/2$ (orange tone) odd
in field. This separation is exposed in their angular dependences,
which follow two different behaviors, one in $\cos^2 \theta$, the
other one in $\cos \theta$, respectively. The solid lines in
FIG.\ref{FIG2}c and in \ref{FIG2}d are a fit by these two
functions.  Comparing the behavior, we observe that the signal
$\Sigma$ is odd in current, while the signal $\Delta$ is even in
current. As noticed in ref \cite{cornelissen15}, these symmetries of
$\Sigma$ and $\Delta$ are the hallmark of respectively SOT
\cite{collet16} and spin Seebeck effects
\cite{uchida10,Jin2015}. Hereafter, we shall use the fit of the whole
angular dependence as a mean to extract precisely the amplitude of
$\Sigma$ and $\Delta$ at $\theta=0$.

FIG.\ref{FIG2}e shows their evolution as a function of current. One
observes that the correspondence between the symmetries of $\Sigma$
and of $\Delta$ with respect to the polarities of $H_0$ and $I$ is
respected (within our measurement accuracy) on the whole current
range. While $\Delta$ approximately follows the parabolic increase of
the Pt temperature (cf. FIG.\ref{FIG1}a), as expected for thermal
effects, the interesting novel feature is the fact that $\Sigma$
deviates from a purely linear transport behavior at large $I$. It is
important also to notice that, when the high/low binding posts of the
current source and voltmeter are biased in the same orientation
(cf. FIG.\ref{FIG1}a), the sign of $(\Sigma \cdot I) < 0$. This is a
signature that the observed non-local voltage is produced by ISHE and
not by leakage electrical currents inside the YIG \cite{Thiery2017a},
although these effects are only expected to occur at much higher
temperatures ($>370$~K\footnote{this corresponds to injecting in the
  Pt wire current densities $>1.0\times
  10^{12}$~A$\cdot$m$^2$}). While in both scenarii the induced
electrical current flows in the same direction in the two parallel Pt
wires, for ISHE, the YIG acts as a source and the potential increases
along the current direction \footnote{this effect is independent of
  the sign of the spin Hall angle}, in contrast, for Ohmic loss, the
YIG acts as a load and the potential drops along the current direction
(\textit{i.e.}  $(\Sigma \cdot I) > 0$ cf. FIG2b in
\cite{Thiery2017a}). We should though add that selecting the component
of the non-local voltage that is $\theta$-dependent is another
effective mean to eliminate Ohmic contribution, since the later are
independent of the in-plane orientation of $H_0$. We have repeated
this measurement on other devices either on the same film or on
different LPE YIG films of similar thickness. On all the devices, we
observe an up-turn of $\Sigma$ at the same current density. While the
sign of $(\Sigma \cdot I)$ is always negative, this is not the case
for the sign for $(\Delta \cdot I)$, which depends on the film quality
(the same $(\Delta \cdot I)$ sign is observed for all devices on the
same film but could change depending on the film quality). Further
progress on the later issue requires a better understanding of the
different phenomena contributing to thermal effects ($\Delta$-signal)
and the means to separate them. In the following, we shall concentrate
exclusively on the non-linear behavior of $\Sigma$ which measures the
number of magnons created by SOT relatively to the number of magnons
annhilated by SOT while being immune to effects caused by Joule
heating.

\begin{figure}
  \includegraphics[width=8.5cm]{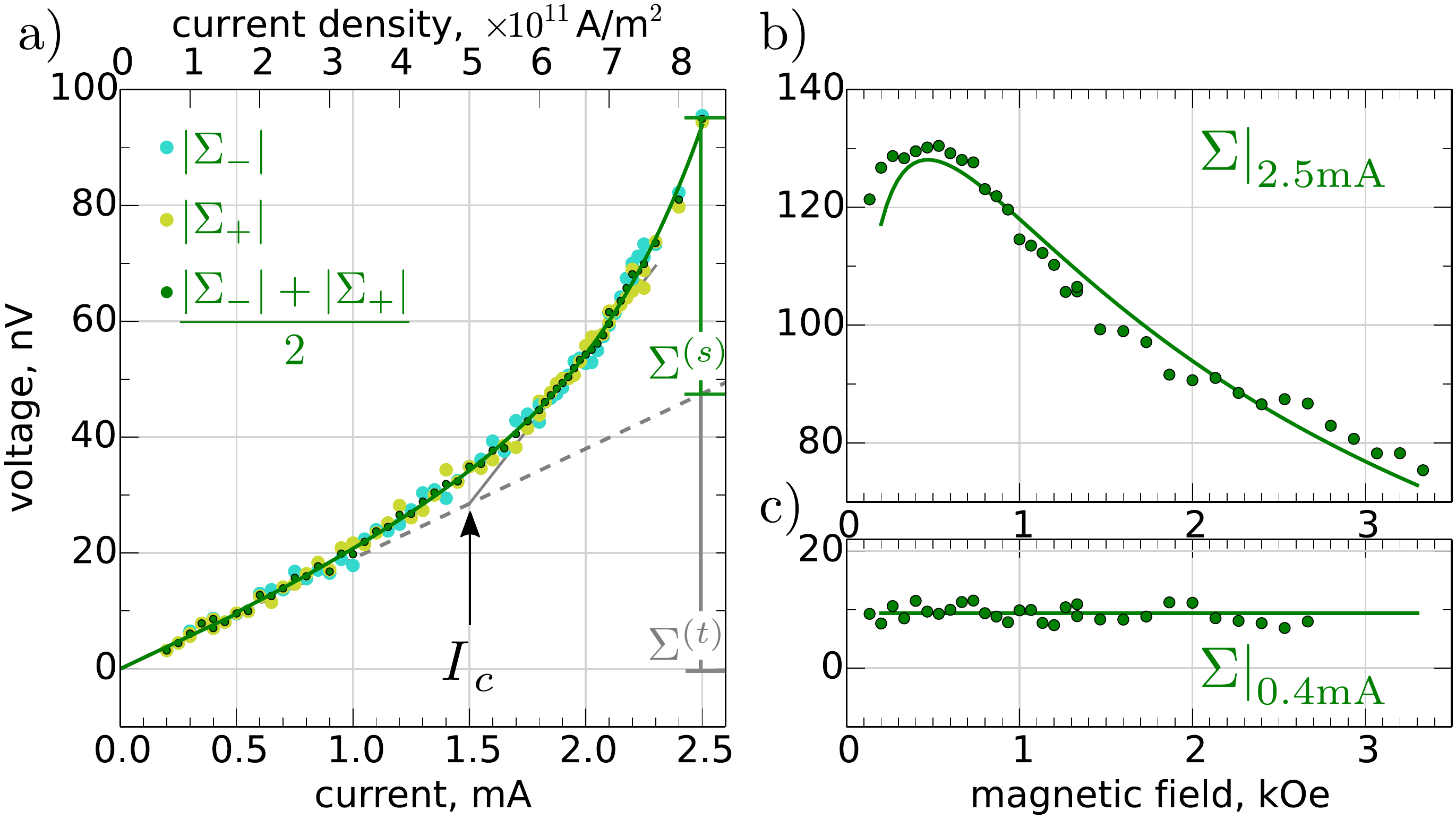}
  \caption{(Color online) Current dependence of the absolute sum
    signal $\Sigma$ averaged over the two current polarities. The
    dashed line is a linear fit of the low current regime and the
    arrow marks the onset at which the spin transport deviates from
    linear behavior. Spin transport excited by SOT can be separated in
    two independent channels: a linear contribution $\Sigma^{(t)}$
    taken on by magnons transporting thermal heat and $\Sigma^{(s)}$,
    an additional magnons' conduction channel that emerges above
    $I_c$. Variation of $\Sigma$ as a function of magnetic field for
    two different current intensities b) above and c) bellow $I_c$.}
  \label{FIG3}
\end{figure}

Using open dots, we have plot in FIG.\ref{FIG3}a both the variation of
$|\Sigma_-|$ and $|\Sigma_+|$ as a function of the current
intensity. Both data set show clearly the emergence of a new spin
transport channel at large current densities, as evidence by the
deviation from a purely linear conduction regime. Since both
quantities $|\Sigma_\mp|$ follow the same behavior on the whole
current range, for the sake of simplicity we shall call simply
$\Sigma$ (dark green) their averaged. At low current, the SOT signal
follows first a linear behavior $\Sigma^{(t)}$ which has been shown to
be dominated by thermal magnons transport \cite{cornelissen15}
\footnote{The non-local linear resistance between the two Pt wires is
  $\Sigma^{(t)}/I=0.019$~m$\Omega$.}. We shall define $\Sigma^{(s)}$
the additional conduction contribution, \textit{i.e.} the deviation
from the extrapolated linear behavior.

Quite remarkably the enhancement of the conductance due to
$\Sigma^{(s)}$ occurs very gradually. We emphasize that such a low
rise is very different from the sudden surge of coherent magnons
observed at the critical thresold in confined geometries
\cite{slavin09,hamadeh12}. Fitting a straight line through the low
current regime, $\Sigma_{I\in [0,0.9]\text{mA}}$, and the high current
deviation, $\Sigma_{I\in [1.6,2.3]\text{mA}}$, the intersection
provides an estimation of the onset current of this new conduction
channel, $I_c\approx 1.5$~mA, which corresponds to a current density
$J_c \approx 5 \times 10^{11}$A/m$^2$. This value is very close to the
threshold current for damping compensation of coherent modes observed
at the same applied field ($H_0=2$~kOe) in micron-sized disks
\cite{collet16} and stripes \cite{evelt16} with similar
characteristics.

More insight about the nature of the magnons excited above $I_c$ can
be obtained by studying the field dependence of $\Sigma$
\cite{Cornelissen2016}. The results are shown in FIG.\ref{FIG3}b and
\ref{FIG3}c for two values of the current $I=0.4$ and 2.5~mA,
respectively below and above $I_c$. While in the field range explored,
the signal is almost independent of $H_0$ when $I<I_c$, it becomes
strongly field dependent when $I>I_c$.  This different behaviors are
consistent with assigning the spin transport to thermal magnons below
$I_c$ and mainly to subthermal magnons above $I_c$. In the former
case, the magnons' energy is of the order of the exchange energy,
which is much larger than the Zeeman energy, while in the latter case,
because of their long wavelength, their energy is of the order of the
magnetostatic energy. In consequence, $\Sigma$ is expected to increase
with decreasing field at fixed $I$, because of the associated decrease
of $I_c$. The behavior scales well with the reduced quantity
$I/I_c$. This is shown by the solid line in FIG.\ref{FIG3}b, which
displays the expected field dependence of $1/I_c(H_0)$
\cite{collet16}) where $I_c \propto (\omega_H+\omega_M/2) \left(
  \alpha + \gamma {\Delta H_0}/({2 \omega_K}) \right)$, where
$\omega_H=\gamma H_0$ and $\omega_M=4 \pi \gamma M_s$, $\gamma$ being
the gyromagnetic ratio, and $\omega_K= \sqrt{\omega_H
  (\omega_H+\omega_M)}$ is the Kittel's law. We have used here the
amount of inhomogeneous broadening, $\Delta H_0=1.5$G (probably
position dependent), as an adjustable parameter, while the value of
the other parameters are those extracted from Table.1.

\begin{figure}
  \includegraphics[width=8.5cm]{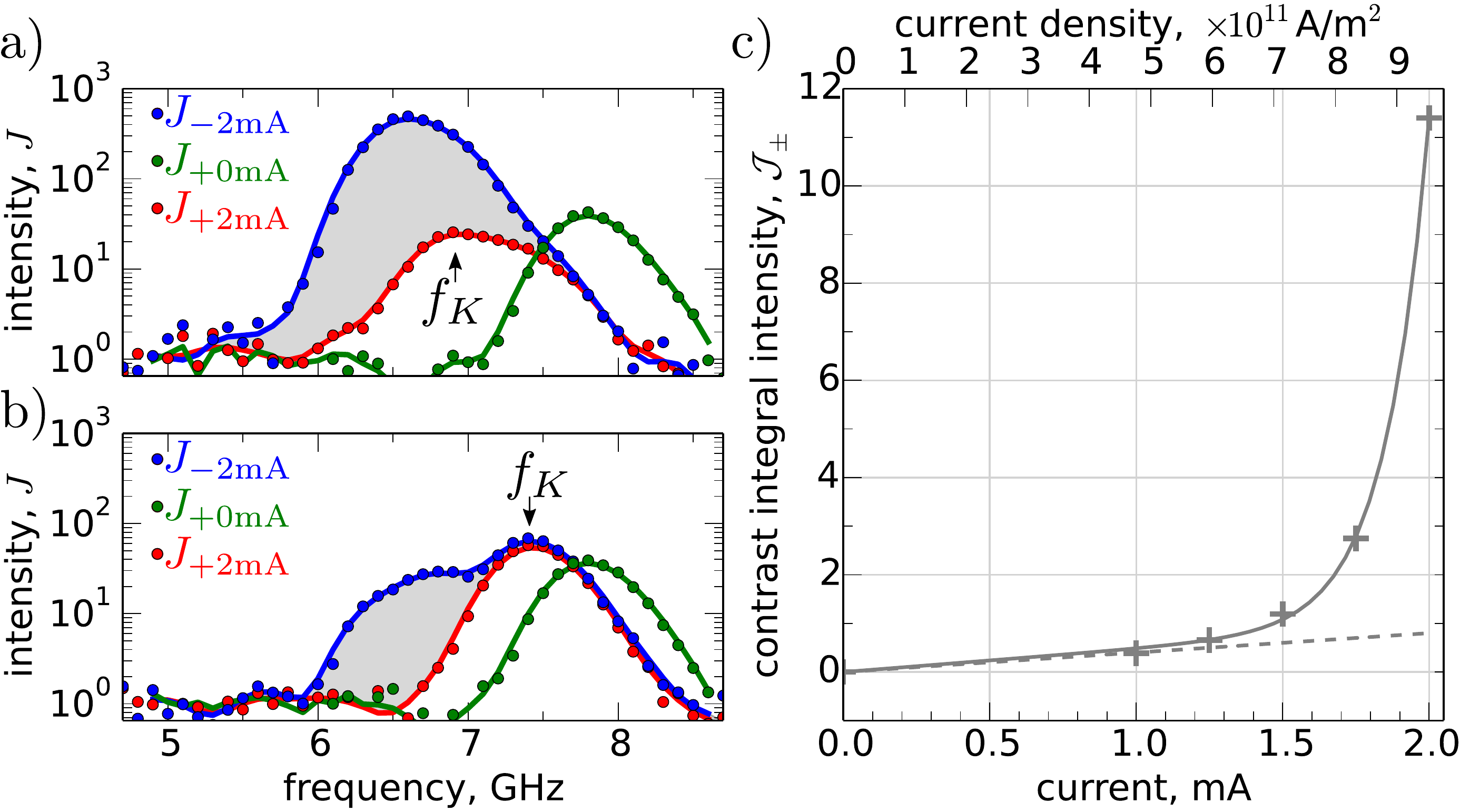}
  \caption{(Color online) Micro-BLS studies of the subthermal magnons
    spectrum at $H_0=+2$~kOe a) underneath the injector and b)
    underneath the detector 0.7$\mu$m away. The left column show the
    spectral distribution of the BLS intensity $J$ measured at
    $I=\pm2$mA. The arrows indicate the Kittel frequency. The
    difference $J_\pm$ (grey area) indicate the spectral distribution
    of the magnons excited by SOT relative to the ones
    annihilated. The panel c) plots the current evolution of the
    integrated intensity ${\cal J}_\pm$ underneath the injector.}
  \label{FIG4}
\end{figure}

The above interpretation has been checked by preforming microfocus
Brillouin light scattering ($\mu$-BLS) in the sub-thermal energy
range. For this measurement, we have used a second series
\footnote{this second series has been used to investigate the
  electrical properties of YIG thin films at high temperature
  \cite{Thiery2017a}} of non-local devices, where the Pt thickness has
been reduced to 7~nm (thus comparison of the results between the 2
series should be done by juxtaposing data obtained with indentical
current densities, cf. upper scale). FIG.\ref{FIG4}a and \ref{FIG4}b
show on a logarithmic scale the spectral distribution of the BLS
intensity, $J$, a) underneath the injector and b) underneath the
detector, which are here separated by $d=0.7\mu$m. The distribution is
measured at $I=\pm 2$~mA (\textit{i.e.} $9.5 \times 10^{11}$A/m$^2$)
while the field is set to $H_0=+2$~kOe and $\theta=0^\circ$. In both
cases, an enhancement of the subthermal magnons population is observed
when $(I \cdot H_0) <0$ (blue), which corresponds to the configuration
where the SOT compensates the damping (cf convention in
FIG\ref{FIG1}a). The measurement for the opposite case $(I \cdot H_0)
>0$ (red) provides a reference about the out-of-equilibrium state
produced by Joule heating. The maximum intensity of the red curve
indicates the resonance frequency of the Kittel mode, $\omega_K/2 \pi$
at the corresponding temperature. This is because the $\mu$-BLS
response function is centered around the long wavelength
magnons. Indeed, the detected signal decreases once the magnon
wavelength is smaller than the spot size (approximately 0.4 $\mu$m:
diffraction limited). In order to isolate the contribution produced by
SOT, we substract the spectral contribution measured at $+I$ to the
one measured at $-I$ (grey shaded area). This allows to cancel out the
spectral deformation produced by Joule heating but, as for the
$\Sigma$-signal, this only measures the enhancement of the magnons
created by SOT relative to the magnons annihilated by SOT. One can
clearly see on the shaded data that SOT enhances the magnons
population in a spectral window between the Kittel frequency and the
bottom of the magnon manifold. Next, we have plotted in FIG\ref{FIG4}c
how the spectral integration of this differential signal ${\cal
  J}_\pm=\int J_\pm d\omega$ varies as a function of the current
amplitude underneath the injector. One observes a regime of linear
rise at small current, followed by a growth above $J_c \approx 5
\times 10^{11}$A/m$^2$ in a similar fashion as the one reported in
FIG\ref{FIG3}a. The $\mu$-BLS experiment thus provides a direct
evidence that an additional spin conduction channel has indeed emerged
in the GHz frequency range (subthermal) at large current when SOT is
in the range to compensate the damping. It also shows that the magnons
newly created are spread at the bottom of the magnon manifold.

In summary, we have shown that while at low values of the spin current
the main contribution to the spin conductance comes from thermal
magnons, the subthermal magnons mainly determine the magnon transport
at high values of the spin current, comparable to the critical
magnitude at which damping compensation of coherent magnons takes
place. We believe that our current findings are not only important
from the from the fundamental point of view, but might be also useful
for future applications. While transport of thermal magnons are
difficult to control due to their relatively high energies, the
subthermal magnons could be efficiently controlled by variation of
relatively weak magnetic fields.

\begin{acknowledgments}

  This research was supported in part by the CEA program NanoScience
  (project MAFEYT), by the priority program SPP1538 Spin Caloric
  Transport (SpinCaT) of the DFG and by the program Megagrant
  14.Z50.31.0025 of the Russian ministry of Education and Science.The
  work at Oakland University was supported by the Grants
  Nos. EFMA-1641989 and ECCS-1708982 from the NSF of the USA, by the
  CND, NRI and by DARPA. VVN acknowledges fellowship from the
  emergence strategic program of UGA, and Russian competitive growth
  program. We thank G. Zhand, T. van Pham, A. Brenac for their help in
  the fabrication of the lateral devices.

\end{acknowledgments}


%

\end{document}